# Mining Top-$K$ Frequent Itemsets Through Progressive Sampling

Andrea Pietracaprina[1,*], Matteo Riondato[2,**], Eli Upfal[2,***], and Fabio Vandin[2,†]

[1] Dipartimento di Ingegneria dell'Informazione, Università di Padova, Padova, Italy.
capri@dei.unipd.it
[2] Department of Computer Science, Brown University, Providence RI, USA.
{matteo,eli,vandinfa}@cs.brown.edu

**Abstract.** We study the use of sampling for efficiently mining the top-$K$ frequent itemsets of cardinality at most $w$. To this purpose, we define an approximation to the top-$K$ frequent itemsets to be a family of itemsets which includes (resp., excludes) all very frequent (resp., very infrequent) itemsets, together with an estimate of these itemsets' frequencies with a bounded error. Our first result is an upper bound on the sample size which guarantees that the top-$K$ frequent itemsets mined from a random sample of that size approximate the actual top-$K$ frequent itemsets, with probability larger than a specified value. We show that the upper bound is asymptotically tight when $w$ is constant.

Our main algorithmic contribution is a progressive sampling approach, combined with suitable stopping conditions, which on appropriate inputs is able to extract approximate top-$K$ frequent itemsets from samples whose sizes are smaller than the general upper bound. In order to test the stopping conditions, this approach maintains the frequency of all itemsets encountered, which is practical only for small $w$. However, we show how this problem can be mitigated by using a variation of Bloom filters.

A number of experiments conducted on both synthetic and real benchmark datasets show that using samples substantially smaller than the original dataset (i.e., of size defined by the upper bound or reached through the progressive sampling approach) enable to approximate the actual top-$K$ frequent itemsets with accuracy much higher than what analytically proved.

## 1 Introduction

For a dataset $\mathcal{D}$ of transactions over an alphabet of items $\mathcal{I}$, the *top-$K$ frequent itemsets* is the family of subsets of items, dubbed *itemsets*, which occur in $\mathcal{D}$ with frequency greater than or equal to the frequency of the $K$-th most frequent one. The extraction of the top-$K$ frequent itemsets is a fundamental primitive which arises directly in applications from many different domains (e.g., data mining, networking, and databases), and it is also regarded as a convenient alternative to the classical data mining problem of computing all itemsets with frequency above a fixed threshold, since the parameter $K$ allows, in practice, for a better control on the output size.

[*] Support provided, in part, by MIUR of Italy under project AlgoDEEP, and by the University of Padova under the Strategic Project STPD08JA32 and Project CPDA099949/09.
[**] Contact Author. Supported in part by NSF award IIS-0905553.
[***] Supported in part by NSF award IIS-0905553.
[†] Supported in part by the "Ing. Aldo Gini" Foundation, Padova, Italy.

For large datasets the mining of top-$K$ frequent itemsets becomes computationally challenging. In particular, exact algorithms must scan the entire dataset more than once and if the dataset does not fit completely in main memory, as is typically the case, disk accesses may slow down the computation to a point where it becomes impractical. In this paper we consider the use of sampling for efficiently mining a suitable approximation to the top-$K$ frequent itemsets from large datasets, and investigate the trade-offs between the sample size and the accuracy of the approximation.

### 1.1 Previous work

A few attempts have been made to devise efficient exact algorithms for mining top-$K$ frequent itemsets from static datasets, restricting the mining task to closed itemsets (i.e., itemsets that have no superset with the same frequency) [14, 11]. However, these algorithms exhibit limited scalability and are unable to efficiently handle extremely large inputs.

A number of works in the literature explored the use of sampling in the context of the classical frequent itemsets mining problem where all itemsets with frequency above a fixed threshold are required [16, 6, 12, 5, 10, 2]. The idea is to extract frequent itemsets from a sample of transactions randomly drawn from the input dataset, where the size of the sample is chosen, possibly much smaller than the original dataset size, so that the frequent itemsets with respect to the sample represent a good approximation to the actual frequent itemsets. Upper bounds on the sample size which guarantee to estimate the frequency of an itemset with a given accuracy and a given confidence are presented in [16, 6]. However, these estimates are derived for the individual itemsets identified by the algorithm and do not bound the probability of not discovering other itemsets of equal or larger frequencies. Ensuring with high probability that all itemsets sought by the mining task be discovered is the main challenge in applying sampling to mining.

In [12] the author proposes an algorithm that, by mining a random sample of the dataset, builds a candidate set of frequent itemsets which contains all the frequent itemsets with a probability that depends on the sample size. However, the sample does not guarantee that all itemsets in the candidate set are frequent. Nevertheless, the set of candidates allows the algorithm to efficiently identify the set of frequent itemsets with at most two passes on the entire dataset.

The use of progressive sampling for mining frequent itemsets has been studied in [5, 10]. Progressive sampling entails analyzing increasingly larger samples until the observed improvement of a certain measure of the accuracy of the sample with respect to the mining task falls below a specified threshold. An empirical two-phase sampling method was devised in [2] which first uses a large sample to estimate the item frequencies accurately, and then uses this information to build, progressively, a suitable sample for mining the frequent itemsets. We remark that the aforementioned works do not assess analytically the relation between the sample size and the accuracy of the approximation to the frequent itemsets gathered from the sample.

The extraction of top-$K$ frequent itemsets through sampling is a more challenging task, since the corresponding minimum frequency threshold is not known in advance. In a recent work [13] sampling is used for extracting, for a fixed parameter $0 < \varepsilon < 1$, an approximation of the top-$K$ frequent items from a sequence of items, which contains no item whose actual frequency is less than $f_K - \varepsilon$, where $f_K$ is the actual frequency of the $K$-th most frequent item. However, to achieve this result the sample size is defined as a function of $f_K$, which is unknown. The authors propose an empirical sequential method to estimate the right sample size. Moreover, the results cannot be directly extended to the mining of top-$K$ frequent item(set)s from datasets of transactions.

A number of recent papers considered the problem of finding (top-$K$) frequent items/itemsets from data streams [4, 7, 1, 8, 15, 3]. In the streaming context, the entire dataset is scanned (at least) once and sampling is employed to maintain summary data structures, with small memory footprints, which provide approximate solutions to the mining task. The accuracy of the approximation is related to the size of the data structures rather than the actual number of elements sampled from the input dataset. In this respect, the objective of these works is slightly different from the one pursued in this paper.

### 1.2 Our contribution

We present novel results on the effectiveness of sampling for mining top-$K$ frequent itemsets from a transactional dataset. In analogy to [13], we define an $\varepsilon$-approximation to the top-$K$ frequent itemsets to be a family of itemsets, together with their estimated frequencies, which excludes all itemsets whose actual frequency is less than $f_K - \varepsilon$ and includes all itemsets whose actual frequency is at least $f_K + \varepsilon$, where $f_K$ is the actual frequency of the $K$-th most frequent itemset. (Note that the latter requirement was not imposed in [13].) Moreover, the estimated frequency of each itemset in the $\varepsilon$-approximation must be at most an additive factor $\varepsilon$ away from the actual one. We also impose a bound $w$ on the size of the frequent itemsets to be discovered

We prove an upper bound $t$ on the sample size which guarantees, with probability at least $1 - \delta$, that an $\varepsilon$-approximation to the top-$K$ frequent itemsets of size up to $w$ is discovered. The bound is a function of the parameters $K$, $\varepsilon$, $\delta$, and of the total number of itemsets of size at most $w$, but it does not depend on the number of transactions in the dataset or on $f_K$. We also show that for $w \in O(1)$ the upper bound $t$ is tight, within constant factors, by arguing the existence of a dataset for which a random sample of size $o(t)$ would not provide the required $\varepsilon$-approximation with sufficiently high probability.

While the above bounds are tight for worst-case datasets, an adaptive approach may be able to compute an $\varepsilon$-approximation to top-$K$ frequent itemsets with smaller samples in some cases. To this purpose, we devise a progressive sampling approach which extracts the top-$K$ frequent itemsets from increasingly larger samples until suitable stopping conditions are met or the upper bound $t$ is hit. A straightforward implementation of our approach requires that in order to test the stopping conditions, the frequencies of all itemsets encountered be maintained, which is practical only if the upper bound $w$ on the itemset size is small. However, we show how this problem can be mitigated by using count-min filters, which are a variation of Bloom filters.

A number of experiments conducted on both synthetic and real benchmark datasets show that the family of top-$K$ frequent itemsets mined either from a random sample of $t$ transactions or from the smaller samples built with the progressive sampling approach approximates the actual top-$K$ frequent itemsets with an accuracy and confidence much higher than what analytically proved. This provides further evidence of the effectiveness of the sampling approach for the mining task under consideration. Moreover, we show analytically on a specific artificial dataset and experimentally on a benchmark, that the stopping conditions employed by the progressive sampling approach are able to stop the sampling schedule before the sample size $t$ is reached, while maintaining the same accuracy and confidence bounds on the output.

The rest of the paper is organized as follows. Section 2 formally defines the notion of $\varepsilon$-approximation to the top-$K$ frequent itemsets. It also introduces some notation and a technical fact which will be used in the analysis. In Section 3 we present the bound $t$ on the sample size sufficient for performing the mining task with the specified confidence and show that it is tight when itemsets of size at most $w = O(1)$ are sought. Section 4 presents the progressive sampling algorithm. A more efficient implementation of the algorithm through count-min filters is outlined in Section 5. Section 6 reports the results of the evaluation of our approach. Section 7 closes the paper with some final remarks.

## 2  Preliminaries

Consider a dataset $\mathcal{D}$ of transactions, where each transaction $\tau$ is a subset of a universe $\mathcal{I}$ of $n$ items. Let $|\tau|$ denote the number of items in transaction $\tau$. Given an itemset $x \in 2^{\mathcal{I}}$, we use $f_{\mathcal{D}}(x)$ (resp., $\sigma_{\mathcal{D}}(x)$) to denote its *frequency* (resp., *support*) in $\mathcal{D}$, namely, the fraction (resp., number) of transactions containing $x$. We consider only itemsets of size at most $w$, and denote with $U(\mathcal{I}, w)$ the complete family of these itemsets. We define $m = |U(\mathcal{I}, w)| = \sum_{i=1}^{w} \binom{n}{i}$.

For convenience, we assume a fixed *canonical ordering* of the itemsets in $U(\mathcal{I}, w)$ by decreasing frequency in $\mathcal{D}$, with ties broken arbitrarily, and label the itemsets $x_1, x_2, \cdots, x_m$ according to this ordering. For a given $K$, with $1 \leq K \leq m$, we denote $f_{\mathcal{D}}^{(K)} = f_{\mathcal{D}}(x_K)$, and define the set of top-$K$ frequent itemsets of size at most $w$ (with their respective frequencies) as

$$\text{TOPK}(\mathcal{D}, \mathcal{I}, K, w) = \left\{ (x, f_{\mathcal{D}}(x)) \, : \, x \in U(\mathcal{I}, w) \, , \, f_{\mathcal{D}}(x) \geq f_{\mathcal{D}}^{(K)} \right\} \, .$$

In this work we aim at efficiently mining the following approximation to the set TOPK($\mathcal{D}, \mathcal{I}, K, w$).

**Definition 1.** *Let $\varepsilon \in (0,1)$ be a real-valued parameter. An $\varepsilon$-approximation to TOPK($\mathcal{D}, \mathcal{I}, K, w$) is a set $W$ of $K$ or more ordered pairs $(x, f)$ such that $x \in U(\mathcal{I}, w)$, $f \in [0,1]$, and for which the following properties hold:*

**P1:** *for each $(x, f) \in W$, $f_\mathcal{D}(x) \geq f_\mathcal{D}^{(K)} - \varepsilon$;*
**P2:** *for each $(x, f) \notin W$, $f_\mathcal{D}(x) < f_\mathcal{D}^{(K)} + \varepsilon$;*
**P3:** *for each $(x, f) \in W$, $|f - f_\mathcal{D}(x)| \leq \varepsilon$.*

Given a sample $\mathcal{S}$, we will use $f_\mathcal{S}(x)$ to denote the frequency of an itemset $x$ in $\mathcal{S}$, and $\sigma_\mathcal{S}(x)$ to denote the support of $x$ in $\mathcal{S}$. The following fact will be used later in the analysis. Due to space constrains, its proof is omitted.

**Fact 1** *Consider a sample $\mathcal{S}$ of $t > 0$ transactions drawn at random with replacement from $\mathcal{D}$. Let $x \in 2^\mathcal{I}$ be an arbitrary itemset. For fixed $\varepsilon \in (0,1)$ and for any itemset $y \in 2^\mathcal{I}$ such that $f_\mathcal{D}(x) \geq f_\mathcal{D}(y) + \varepsilon$ we have:*

$$\Pr(f_\mathcal{S}(y) > f_\mathcal{S}(x)) \leq e^{-\frac{\varepsilon^2}{2}t} \text{ and} \qquad (1)$$

$$\Pr(|f_\mathcal{S}(x) - f_\mathcal{D}(x)| \geq \varepsilon) \leq 2e^{-\frac{\varepsilon^2}{2}t} \ . \qquad (2)$$

## 3   Upper bound on the sample size

Consider a sample $\mathcal{S}$ of $t$ transactions drawn at random with replacement from $\mathcal{D}$. Theorem 1 below shows that if $t$ is large enough then the set of top-$K$ frequent itemsets from $\mathcal{S}$, with their respective frequencies in the sample, yields an $\varepsilon$-approximation to TOPK($\mathcal{D}, \mathcal{I}, K, w$), with a certain probability.

**Theorem 1.** *For fixed $\varepsilon, \delta \in (0,1)$, let $\mathcal{S}$ be a sample of*

$$t = \frac{2}{\varepsilon^2} \ln \frac{2m + K(m - K)}{\delta}$$

*transactions drawn at random with replacement from $\mathcal{D}$. Then, $W = $ TOPK($\mathcal{S}, \mathcal{I}, K, w$) is an $\varepsilon$-approximation to TOPK($\mathcal{D}, \mathcal{I}, K, w$) with probability at least $1 - \delta$.*

*Proof.* Consider the $m$ itemsets of $U(\mathcal{I}, w)$, namely $x_1, x_2, \cdots, x_m$, indexed according to the canonical ordering introduced before, and define the following property **Q**: "for every $i, j$, with $1 \leq i \leq K < j \leq m$, such that $f_\mathcal{D}(x_i) \geq f_\mathcal{D}(x_j) + \varepsilon$ we have $f_\mathcal{S}(x_i) > f_\mathcal{S}(x_j)$". We first show that if property **Q** holds then properties **P1** and **P2** of Definition 1 hold for $W$. Assume that **Q** holds. Then, the frequency in $\mathcal{S}$ of any $x_i$, with $1 \leq i \leq K$, is larger than the frequency in $\mathcal{S}$ of any itemset $x_j$ with $f_\mathcal{D}(x_j) < f_\mathcal{D}^{(K)} - \varepsilon$, and this implies Property **P1**. As for **P2**, consider a pair $(x, f_\mathcal{S}(x)) \notin W$ and suppose, by contradiction, that $f_\mathcal{D}(x) \geq f_\mathcal{D}^{(K)} + \varepsilon$, that is $x = x_i$, for some index $i$ with $1 \leq i < K$. If this is the case, since $W$ contains at least $K$ pairs, it must contain a pair $(x_j, f_\mathcal{S}(x_j))$ with $j > K$ and $f_\mathcal{S}(x_j) > f_\mathcal{S}(x_i)$. This is impossible because **Q** holds and $f_\mathcal{D}(x_i) \geq f_\mathcal{D}(x_j) + \varepsilon$.

We complete the proof by showing that if the sample size $t$ is chosen as stated, then, with probability at least $1 - \delta$, both **Q** and **P3** (from Definition 1) hold. Consider a pair of itemsets $(x_i, x_j)$ with $1 \leq i \leq K < j \leq m$ such that $f_\mathcal{D}(x_i) \geq f_\mathcal{D}(x_j) + \varepsilon$. By Fact 1 we have $\Pr(f_\mathcal{S}(x_j) > f_\mathcal{S}(x_i)) \leq e^{-\frac{\varepsilon^2}{2}t}$. Also, from Relation (2) we have that for any itemset $x_i$, with $1 \leq i \leq m$, $\Pr(|f_\mathcal{S}(x_i) - f_\mathcal{D}(x_i)| \geq \varepsilon) \leq 2e^{-\frac{\varepsilon^2}{2}t}$. Note that $K(m-K)$ pairs are involved in **Q** and at most $m$ itemsets are involved in property **P3**. Therefore, by applying an union bound, the probability that **Q** or **P3** do not hold is at most $(2m + K(m-K))e^{-\frac{\varepsilon^2}{2}t} \leq \delta$, and the theorem follows. □

Note that $t$ is independent of the number of transactions in $\mathcal{D}$ and of the frequencies of the itemsets in $\mathcal{D}$.

We now show that for $K, w \in O(1)$ and constant $\varepsilon$, if we fix the confidence parameter $\delta$ suitably small, yet constant, then for a random sample $\mathcal{S}$ of size $t \in o(\ln m) = o((1/\varepsilon^2)\ln(m/\delta))$ the probability that $\text{TOPK}(\mathcal{S}, \mathcal{I}, K, w)$ is an $\varepsilon$-approximation to $\text{TOPK}(\mathcal{D}, \mathcal{I}, K, w)$ can be made smaller than $1 - \delta$ by choosing the number of items and the number of transactions in $\mathcal{D}$ large enough. Consider a universe $\mathcal{I}(\ell)$ of $K + \ell$ items, namely $\mathcal{I}(\ell) = \{y_1, \ldots, y_K\} \cup \{x_1, \ldots, x_\ell\}$, where $K$ is fixed and $\ell$ is a parameter.

**Theorem 2.** *Let $K, w \in O(1)$ and consider an arbitrary constant $\varepsilon \in (0, 1/4)$. Fix a constant $\delta < 1 - (1/2)^K$. Let $t(x)$ be any integer-valued function such that $t(x) \in o(\ln x)$. Then, for large enough $\ell$, there exists a dataset $D^*$ over $\mathcal{I}(\ell)$ such that for a random sample $\mathcal{S}$ of $t(\ell)$ transactions, the probability that $\text{TOPK}(\mathcal{S}, \mathcal{I}, K, w)$ is an $\varepsilon$-approximation to $\text{TOPK}(\mathcal{D}, \mathcal{I}, K, w)$ is less than $1 - \delta$.*

*Proof.* Fix $p_K \in (0, 1)$ such that $p_K > 2\varepsilon$ and $p_K - p_K^2 > \varepsilon$, and let $p_\ell = p_K - 2\varepsilon$. Consider a random dataset $\mathcal{D}$ of $N$ transactions over $\mathcal{I}(\ell)$, where in any transaction each item $y_i$ (resp., $x_j$) is included with probability $p_K$ (resp., $p_\ell$) independently of all other items and all other transactions. Hence, $p_K$ is the expected frequency of each $y_i$, and $p_\ell$ the expected frequency of each $x_j$. Let $\hat{p}$ be the minimum frequency in $\mathcal{D}$ of the items $y_1, \ldots, y_K$. It can be easily shown that, by the choice of $p_K$, if $N$ is large enough the event $F = \text{"TOPK}(\mathcal{D}, \mathcal{I}, K, w) = \{y_1, \ldots, y_K\}$ *and all other itemsets have frequency smaller than* $\hat{p} - \varepsilon$" holds with probability $1 - o(1)$. Thus the only itemsets that can be reported in an $\varepsilon$-approximation are the items $\{y_1, \ldots, y_K\}$. For a sample $\mathcal{S}$, define the events $E_1 = $ "*at least one of $x_1, \ldots, x_\ell$ appears in $\mathcal{S}$ with frequency $\geq p_K$*", and $E_2 = $ "*at least one of $y_1, \ldots, y_K$ appears in $\mathcal{S}$ with frequency $\leq p_K$*". When $F$, $E_1$ and $E_2$ occur, $\text{TOPK}(\mathcal{S}, \mathcal{I}, K, w)$ does not satisfy property **P1**.

In what follows we prove that if $\mathcal{S}$ is a random sample of $t(\ell) = o(\ln \ell) = o(\ln m)$ transactions from a dataset $\mathcal{D}$ built as above, then $\Pr(F \cap E_1 \cap E_2) \geq 1 - o(1) - (1/2)^K$. This implies that $\Pr(E_1 \cap E_2 | F) \geq 1 - o(1) - (1/2)^K$, and thus there must exist a dataset $D^*$ for which the event $F$ holds and such that if $\mathcal{S}$ is a random sample of $t(\ell)$ transactions then $\Pr(\text{TOPK}(\mathcal{S}, \mathcal{I}, K, w)$ is an $\varepsilon$-approximation to $\text{TOPK}(D^*, \mathcal{I}, K, w)) \leq (1/2)^K + o(1)$. Hence, this probability can be made smaller than $1 - \delta$ by choosing $N$ and $\ell$ sufficintly large.

Since we already argued that for $N$ large enough event $F$ occurs with probability at least $1 - o(1)$, it is sufficient to prove that $\Pr(E_1 \cap E_2) \geq 1 - o(1) - (1/2)^K$. We first show that if $\mathcal{S}$ has size $t(\ell)$ then $\Pr(E_1) = 1 - o(1)$. Let $X_i = 1$ if $f_{\mathcal{S}}(x_i) \geq p_K$ and $X_i = 0$ otherwise. Then $X = \sum_{i=1}^{\ell} X_i$ is a random variable counting how many of the items $x_1, \ldots, x_\ell$ appear in $\mathcal{S}$ with frequency at least $p_K$. Thus $\Pr(E_1) = \Pr(X \geq 1)$. We have[3]

$$\Pr(X_i = 1) \geq \binom{t}{tp_K} p_\ell^{tp_K} (1 - p_\ell)^{t - tp_K} \geq \left(\frac{p_\ell}{p_K}\right)^{tp_K} (1 - p_\ell)^{t - tp_K}$$

Let $d > 1$ be a constant such that $1/d = \min\{p_\ell/p_K, 1 - p_\ell\}$. Then $(p_\ell/p_K)^{tp_K}(1 - p_\ell)^{t - tp_K} \geq (1/d)^t$. We then have $\Pr(X = 0) \leq \left(1 - \frac{1}{d^t}\right)^\ell \leq e^{-\ell/d^t}$. If $t \in o(\ln \ell) = o(\ln m)$, we have $\Pr(X = 0) = o(1)$, that proves $\Pr(E_1) = \Pr(X \geq 1) = 1 - \Pr(X = 0) = 1 - o(1)$.

Now we turn our attention to the $K$ items $y_1, \ldots, y_K$. We have

$$\Pr(E_2) = 1 - \prod_{i=1}^{K} \Pr(f_{\mathcal{S}}(y_i) > p_K) \geq 1 - \left(\frac{1}{2}\right)^K$$

where the last inequality follows from the fact that $p_K$ is the expected frequency of $y_i$. Thus, $\Pr(E_1 \cap E_2) \geq 1 - o(1) - (1/2)^K$ and the theorem follows. □

## 4 Algorithm for approximating the top-$K$ frequent itemsets

We now describe an efficient algorithm which discovers an $\varepsilon$-approximation to $\text{TOPK}(\mathcal{D}, \mathcal{I}, K, w)$ by mining progressively larger samples of the dataset $\mathcal{D}$ until the sample size established in Theorem 1 is reached, or a certain stopping condition is met. When the algorithm stops, it returns, as output, the set $\text{TOPK}(\mathcal{S}^*, \mathcal{I}, K, w)$, where $\mathcal{S}^*$ is the last processed sample. For $j \geq 0$, define

$$t_j = \frac{8}{\varepsilon^2}\left(\ln\frac{8K}{\delta} + j\right) .$$

Let also $j_{\max} \geq 0$ be the smallest index such that

$$t_{j_{\max}} \geq \min\left\{|\mathcal{D}|, (2/\varepsilon^2)\ln((2m + K(m - K))/\delta)\right\} .$$

The algorithm performs a sequence of phases. Specifically, in Phase $j$, for $j \geq 0$ and $j < j_{\max}$, the algorithm processes a random sample of $t_j$ transactions.

---
[3] For notational convenience, we replace $t(\ell)$ with $t$ in the formulas.

A different sample schedule can be used, provided that the size of the sample processed at Phase $j$ is at least $t_j$, and the results we present would still hold. In practice it may be more efficient to use a geometrical progression of sample sizes, starting at $t_0$ defined as above.

In Phase $j_{\max}$, if $t_{j_{\max}} \geq |\mathcal{D}|$ the algorithm processes $\mathcal{D}$ to extract TOPK$(\mathcal{D}, \mathcal{I}, K, w)$, otherwise it considers a random sample of $t_{j_{\max}}$ transactions. The algorithm stops when $j = j_{\max}$, or $j < j_{\max}$ and a suitable stopping condition (specified below) holds.

Consider Phase $j$ and let $\mathcal{S}$ be the random sample of size $t_j$ processed in the phase. Define $\sigma_j = K \left(\frac{e}{2}\right)^j$. For $i \geq 0$, define also $s_j(i) = \lfloor (2\sigma_j)^{(i+1)^2}/2 \rfloor$, and $S_j(i) = \sum_{\ell=0}^{i} s_j(\ell)$. For notational convenience, we assume $S_j(-1) = 0$ and use $h(j)$ as the largest index such that $S_j(h(j) - 1) + 1 \leq m$. Consider an ordering of the itemsets of $U(\mathcal{I}, w)$ by decreasing frequency w.r.t. $\mathcal{S}$, and let $f_{\mathcal{S}}^{(\ell)}$ denote the frequency in $\mathcal{S}$ of the $\ell$-th itemset of $U(\mathcal{I}, w)$ in this ordering. The *stopping condition for Phase $j$* is

$$f_{\mathcal{S}}^{(K)} - f_{\mathcal{S}}^{(S_j(i-1)+1)} > (i+1)\varepsilon \quad \text{for} \quad 1 \leq i \leq h(j) \ . \tag{3}$$

A pseudocode for the algorithm is given in Algorithm 1, where the function `StoppingConditionIsSatisfied` checks whether the above stopping condition holds. The efficient implementation of this function is discussed in Section 5.

---

**Algorithm 1:** Pseudocode of the algorithm

  **input**  : dataset $\mathcal{D}$, integers $K$, $w$, reals $\varepsilon$, $\delta : 0 < \varepsilon, \delta < 1$.
  **output**: A collection of ordered pairs $(x, f)$ which is an $\varepsilon$-approximation to
  TOPK$(\mathcal{D}, \mathcal{I}, K, w)$ with probability at least $1 - \delta$.

1 $m \leftarrow |U(\mathcal{I}, w)|$; $bound \leftarrow \min\left\{|\mathcal{D}|, \frac{2}{\varepsilon^2} \ln \frac{(2m + K(m-K))}{\delta}\right\}$
2 $j \leftarrow 1$; $j_{\max} \leftarrow \arg\min \left\{ z : \frac{8}{\varepsilon^2} \left(\ln \frac{8K}{\delta} + z\right) > bound \right\}$
3 **while** $j < j_{\max}$ **do**
4 $\quad$ $t_j \leftarrow \frac{8}{\varepsilon^2} \left(\ln \frac{8K}{\delta} + j\right)$
5 $\quad$ $\mathcal{S} \leftarrow$ random sample of size $t_j$ from $\mathcal{D}$
6 $\quad$ **if** `StoppingConditionIsSatisfied` **then return** TOPK$(\mathcal{S}, \mathcal{I}, K, w)$
$\quad$ $\quad$ $j \leftarrow j + 1$
7 **end**
8 **if** $bound < |\mathcal{D}|$ **then** $\mathcal{S} \leftarrow$ a random sample of size $bound$ from $\mathcal{D}$
9 **else** $\mathcal{S} \leftarrow \mathcal{D}$
10 **return** TOPK$(\mathcal{S}, \mathcal{I}, K, w)$

---

We now show that, with probability at least $1 - \delta$, the set returned by the above algorithm is an $\varepsilon$-approximation to TOPK$(\mathcal{D}, \mathcal{I}, K, w)$. For Phase $j$ of the algorithm we define $B_j(i)$, with $0 \leq i \leq h(j)$, as the set of $s_j(i)$ itemsets of $U(\mathcal{I}, w)$ whose rank in the canonical ordering (w.r.t. the original dataset $\mathcal{D}$) is in the interval $[S_j(i-1) + 1, S_j(i)]$.

**Lemma 1.** *The following property holds with probability at least $1-\delta$: for every Phase $j$ of the algorithm, for every $0 \leq i \leq h(j)$, and for every itemset $x \in B_j(i)$:*

$$|f_\mathcal{S}(x) - f_\mathcal{D}(x)| < (i+1)\frac{\varepsilon}{2},$$

*where $\mathcal{S}$ is the sample processed in Phase $j$.*

*Proof.* Let us focus on an arbitrary Phase $j$. From Fact 1, Relation (2), we have that for any $x \in B_j(i)$

$$\Pr\left(|f_\mathcal{S}(x) - f_\mathcal{D}(x)| \geq (i+1)\frac{\varepsilon}{2}\right) \leq 2e^{-\varepsilon^2(i+1)^2 t_j/8} .$$

Hence the probability that there exists an itemset $x$ (belonging to any $B_j(i)$) for which the stated bound does not hold is upper bounded by:

$$\sum_{i=0}^{h(j)} s_j(i) 2e^{-\varepsilon^2(i+1)^2 t_j/8} \leq \sum_{i=0}^{h(j)} \left(2\sigma_j e^{-\varepsilon^2 t_j/8}\right)^{(i+1)^2} = \sum_{i=0}^{h(j)} \left(\frac{\delta}{2^{j+2}}\right)^{(i+1)^2} \leq \frac{\delta}{2^{j+1}} .$$

The lemma follows by applying the union bound over all phases (i.e., $j = 0, 1, \ldots$). □

The following theorem establishes the desired probabilistic guarantee on the correctness of the algorithm.

**Theorem 3.** *The algorithm returns an $\varepsilon$-approximation to $\text{TOPK}(\mathcal{D}, \mathcal{I}, K, w)$ with probability at least $1 - \delta$.*

*Proof.* We consider two cases, depending on when the algorithm stops. If the algorithm stops at Phase $j = j_{\max}$, then the output is correct since it coincides with the set $\text{TOPK}(\mathcal{D}, \mathcal{I}, K, w)$, if $t_{j_{\max}} \geq |\mathcal{D}|$, or, otherwise, it is an $\varepsilon$-approximation to $\text{TOPK}(\mathcal{D}, \mathcal{I}, K, w)$ with probability at least $1 - \delta$ as shown by Theorem 1. Suppose instead that the algorithm stops at an earlier phase $j < j_{\max}$ because the stopping condition for Phase $j$ is met, and let $\mathcal{S}$ denote the sample used in this phase. By Lemma 1, for every $0 \leq i \leq h(j)$, and for every itemset $x \in B_j(i)$, we have $|f_\mathcal{S}(x) - f_\mathcal{D}(x)| < (i+1)\frac{\varepsilon}{2}$. Let $W = \text{TOPK}(\mathcal{S}, \mathcal{I}, K, w)$ be the set returned by the algorithm. We now prove that $W$ satisfies properties **P1**, **P2**, and **P3** of Definition 1.

We first show that for each $(x, f_\mathcal{S}(x)) \in W$, we have that $x \in B_j(0)$. By contradiction, assume that for some $(x, f_\mathcal{S}(x)) \in W$, $x \in B_j(i)$, for some $i > 0$. Hence, $f_\mathcal{D}(x) \leq f_\mathcal{D}^{(S_j(i-1)+1)}$ and

$$f_\mathcal{S}^{(K)} \leq f_\mathcal{S}(x) \leq f_\mathcal{D}(x) + (i+1)\frac{\varepsilon}{2} \leq f_\mathcal{D}^{(S_j(i-1)+1)} + (i+1)\frac{\varepsilon}{2} . \quad (4)$$

Observe that all itemsets whose rank in the canonical ordering (w.r.t. $\mathcal{D}$) is not larger than $S_j(i-1) + 1$ belong to sets $B_j(\ell)$ with $\ell \leq i$. By Lemma 1, for each such itemset $z$, we have that

$$f_\mathcal{S}(z) \geq f_\mathcal{D}(z) - (i+1)\frac{\varepsilon}{2} \geq f_\mathcal{D}^{(S_j(i-1)+1)} - (i+1)\frac{\varepsilon}{2} .$$

Hence, since there are $S_j(i-1)+1$ of these itemsets, it follows that

$$f_{\mathcal{S}}^{(S_j(i-1)+1)} \geq f_{\mathcal{D}}^{(S_j(i-1)+1)} - (i+1)\frac{\varepsilon}{2} \ . \tag{5}$$

By combining Equations 4 and 5 we obtain that $f_{\mathcal{S}}^{(K)} - f_{\mathcal{S}}^{(S_j(i-1)+1)} \leq (i+1)\varepsilon$, which contradicts the stopping condition (3). Thus, all itemsets occurring in $W$ belong to $B_j(0)$. This fact, together with the inequality stated in Lemma 1 for the itemsets of $B_j(0)$, establishes Property **P3**.

Now, if we consider any of the first $K$ itemsets of $U(\mathcal{I}, w)$ in the canonical ordering, say $x_\ell$, for some $1 \leq \ell \leq K$, which belongs to $B_j(0)$ by construction, we have that $f_{\mathcal{S}}(x_\ell) \geq f_{\mathcal{D}}(x_\ell) - \frac{\varepsilon}{2} \geq f_{\mathcal{D}}^{(K)} - \frac{\varepsilon}{2}$. Hence, $f_{\mathcal{S}}^{(K)} \geq f_{\mathcal{D}}^{(K)} - \frac{\varepsilon}{2}$. Therefore, for each $(x, f_{\mathcal{S}}(x)) \in W$ we have

$$f_{\mathcal{D}}(x) \geq f_{\mathcal{S}}(x) - \frac{\varepsilon}{2} \geq f_{\mathcal{S}}^{(K)} - \frac{\varepsilon}{2} \geq f_{\mathcal{D}}^{(K)} - \varepsilon,$$

which establishes Property **P1**. Finally, in order to establish Property **P2**, we observe that $W$ must contain a pair $(z, f_{\mathcal{S}}(z))$ such that $f_{\mathcal{D}}(z) \leq f_{\mathcal{D}}^{(K)}$. As argued before, $z \in B_j(0)$, hence

$$f_{\mathcal{S}}^{(K)} \leq f_{\mathcal{S}}(z) \leq f_{\mathcal{D}}(z) + \frac{\varepsilon}{2} \leq f_{\mathcal{D}}^{(K)} + \frac{\varepsilon}{2} \ . \tag{6}$$

Consider an itemset $y \in U(\mathcal{I}, w)$ such that $(y, f_{\mathcal{S}}(y)) \notin W$. If $y \in B_j(i)$ with $i > 0$ then by definition of $B_j(i)$ its real frequency is at most $f_{\mathcal{D}}^{(K)}$, hence it cannot be greater than or equal to $f_{\mathcal{D}}^{(K)} + \varepsilon$. If instead $y \in B_j(0)$ we have

$$f_{\mathcal{D}}(y) \leq f_{\mathcal{S}}(y) + \frac{\varepsilon}{2} < f_{\mathcal{S}}^{(K)} + \frac{\varepsilon}{2} \leq f_{\mathcal{D}}^{(K)} + \varepsilon,$$

where the last inequality follows from Equation 6.) Thus, Property **P2** is established. □

## 5 Efficient implementation with count-min filter

A straightforward implementation of function `StoppingConditionIsSatisfied` presented in Section 4 requires $m = |U(\mathcal{I}, w)|$ counters to store the observed frequencies of all itemsets in order to evaluate the stopping condition (3). We now describe an efficient implementation which uses count-min filters, a variation of Bloom filters, to save space. For a dataset with $O(1)$ transaction size the use of count-min filters reduces the space requirements from $m$ to $O(\log m)$ counters.

A count-min filter $B$ consists of $c$ counters, and uses $k_B$ hash functions. The counters are split into $k_B$ disjoint groups of size $\frac{c}{k_B}$ (we assume that $k_B$ divides $c$ evenly). The $k_B$ hash functions map itemsets into counters, so each hash function $H_i, 1 \leq i \leq k_B$ is a map from the set $U(\mathcal{I}, w)$ to the integers in the range $[(i-1)c/k_B, ic/k_B - 1]$. A more detailed description of count-min filters and their properties can be found in [9, Section 13.4]. Given a sample

$\mathcal{S}$, we use a count-min filter $B$ to approximatethe frequencies of itemsets in $\mathcal{S}$. Initially, all counters are set to 0, then, for each transaction $\tau \in \mathcal{S}$ and each itemset $x \in U(\mathcal{I}, w)$ appearing in $\tau$, we increment by one the $k_B$ counters $H_i(x)$ associated with $x$.

We now introduce some definitions and some results on count-min filters which we will use later in the analysis. The *count-min support* of an itemset $x$ is the value of the minimum of the $k_B$ counters associated with $x$ in $B$, and is denoted with $\sigma_B(x)$. The *count-min frequency* of $x$ is $f_B(x) = \frac{\sigma_B(x)}{|\mathcal{S}|}$. (In the notation for count-min support and count-min frequency we omit any reference to $\mathcal{S}$ because the set of transactions on which the count-min filter is built will be clear from the contest.) We denote the sum of the number of itemsets from $U(\mathcal{I}, w)$ in the transactions of $\mathcal{S}$ as $C_\mathcal{S} = \sum_{\tau \in \mathcal{S}} \sum_{i=0}^{w} \binom{|\tau|}{i}$.

The following theorem (proof omitted due to space constraints) shows that we can obtain a good approximation of the frequencies of the itemsets using a count-min filter.

**Theorem 4.** *Given $\delta_B > 0$, $\varepsilon_B > 0$, and a sample $\mathcal{S}$, let $\varepsilon_C = \frac{\varepsilon_B |\mathcal{S}|}{C_\mathcal{S}}$ and $\delta_c = \delta_B/m$. If $B$ is a count-min filter of parameters $k_B = \left\lceil \ln \frac{1}{\delta_c} \right\rceil$ and $c = \left\lceil \ln \frac{1}{\delta_c} \right\rceil \cdot \left\lceil \frac{e}{\varepsilon_C} \right\rceil$, then $\Pr(\exists x | f_B(x) \geq f_\mathcal{S}(x) + \varepsilon_B) \leq \delta_B$.*

While the count-min filter is useful in reducing the space required to approximate the frequencies of itemsets in the sample, it is not trivial to check the stopping condition without explicitly querying the filter for the count-min frequency of *every* itemset in $U(\mathcal{I}, w)$ (not only of those that appear in the sample), an operation that can be computationally too expensive. Our algorithm will make use of an approximation of the distribution of the frequencies in the sample of the itemsets, built using only the min-count frequencies that appear in the min-count filter, without generating all the itemsets. The algorithm uses the same parameters $\mathcal{D}$, $K$, $\varepsilon$, and $\delta$ as the algorithm of Section 4.

Let $\delta_1, \delta_2 > 0$ such that $(1-\delta_1)(1-\delta_2) = 1-\delta$. We define $t_j$ similarly to Section 4, using $\delta_1$ instead of $\delta$. Let $j_{\max}$, $\sigma_K$, $s_j(i)$, $S_j(i)$, and $h(j)$ be defined as in the algorithm of Section 4. The algorithm performs a sequence of phases and stops when $j = j_{\max}$, or $j < j_{\max}$ and a suitable stopping condition (Equation (7) specified below) holds.

Let $\mathcal{S}$ be the sample analyzed by the algorithm at phase $j$. At each phase the algorithm will use a count-min filter $B$ with parameters $c, k_B$ tuned so that $\Pr(\exists x | f_B(x) \geq f_\mathcal{S}(x) + \varepsilon_B) \leq \delta_2$ (see Theorem 4). Note that $\varepsilon_B$ is not defined by the user. First, the algorithm obtains $\mathrm{TOPK}(\mathcal{S}, \mathcal{I}, K, w)$. Then, we scan the sample and populate the min-count filter $B$ as described before. With a second scan of the sample, the algorithm computes an approximation $\hat{f}$ to the distribution of frequencies of itemsets in the samples, so that for all $j$, if $\hat{f}^{(j)}$ is the frequency of the $j$-th most frequent itemset using this approximation, $\hat{f}^{(j)} \geq f_\mathcal{S}^{(j)}$ holds. For each itemset $x \in U(\mathcal{I}, w)$ appearing in $\mathcal{S}$, let $c_x$ be the counter of $B$ with minimum value among those associated to $x$, and let, for each counter $\ell$ of $B$, $\mathcal{I}_\ell = \{x \in U(\mathcal{I}, w) : c_x = \ell\}$. The approximation $\hat{f}$ is computed as follows: for each counter $\ell$ of $B$, another counter $s_\ell$ is created, with initial value zero. The algorithm scans $\mathcal{S}$ and, for each transaction $\tau$ in $\mathcal{S}$ and each itemset $x$ of length up to $w$ appearing in $\tau$, the algorithm increases $s_{c_x}$ by one. Once the scan is terminated, the value of $s_\ell$ will be $v_\ell = \sum_{x \in \mathcal{I}_\ell} \sigma_\mathcal{S}(x)$. Since we built $B$ so that, with probability at least $1 - \delta_2$, for each $x \in \mathcal{I}_\ell$ we have $\sigma_\mathcal{S}(x) \geq \sigma_B(x) - \varepsilon_B|\mathcal{S}|$ then, if Theorem 4 holds, the value $r_\ell = \left\lfloor \frac{v_\ell}{\sigma_B(x) - \varepsilon_B|\mathcal{S}|} \right\rfloor$ is an upper bound to $|\mathcal{I}_\ell|$. For each counter $\ell$ of $B$ let $x$ be an itemset in $\mathcal{I}_\ell$, then we define $f_\ell = \frac{\sigma_B(x)}{|\mathcal{S}|}$. To obtain an ordering for the approximate frequencies of all itemsets in $U(\mathcal{I}, w)$, we need to sort only the $p$ frequencies of the $p$ counters in $B$, since in our approximation there will be $r_\ell$ itemsets with frequency $f_\ell$. Let $\hat{f}^{(i)}$ be the frequency at the $i$-th position in this order, for $1 \leq i \leq m$. By definition of $\hat{f}^{(i)}$ and $\mathcal{I}_\ell$, we have that $\hat{f}^{(i)} \geq f_\mathcal{S}^{(i)}$. The stopping condition for phase $j$ is then

$$f_\mathcal{S}^{(K)} - \hat{f}^{(S_j(i-1)+1)} > (i+1)\varepsilon \text{ for } 1 \leq i \leq h(j) \ . \tag{7}$$

Note that the choice of $\varepsilon_B$ influences the stopping condition, since the accuracy of $\hat{f}$ depends on $\varepsilon_B$. When the algorithm stops, it returns the set of top-$K$ frequent itemsets and their respective frequencies with respect to the last processed sample. The following theorem (proof omitted due to space constraints) easily follows from the considerations above.

**Theorem 5.** *The output of the min-count filter based algorithm is an $\varepsilon$-approximation to* $\text{TOPK}(\mathcal{D}, \mathcal{I}, K, w)$ *with probability at least* $1 - \delta$.

## 6 Evaluation

In this section, we provide evidence of the effectiveness of our results. Specifically, in Subsection 6.1 we evaluate experimentally the quality of the approximation of the top-$K$ frequent itemsets obtained by mining small samples with the sizes derived in the previous sections. In Subsection 6.2 we provide both analytical and experimental evidence that the stopping condition used by the algorithm presented in Section 4 is effective for certain datasets.

### 6.1 Evaluation of the quality of the output

We first conduct an analysis of the "quality" of the output set obtained either by mining a sample of a dataset with a size set by the bound presented in Theorem 1, or by running the algorithm of Section 4. We used the real datasets *kosarak* and *webdocs*, and the artificial dataset *T10I4D100K* from the FIMI repository [4] whose main characteristics are synthetized in Figure 1.

Each of these datasets has a different distribution of the frequencies of the items. We used several values for $K$ $(1; 2; 5; 10; 100; 1000)$ and the values 1, 2, and 3 for $w$. In all of our experiments, $\delta$ was fixed to 0.1 and $\varepsilon$ to 0.02. For each combination (dataset, $K$, $w$), we mined $\text{TOPK}(\mathcal{S},\mathcal{I},k,w)$ from 100 random samples $\mathcal{S}$ of size derived from Theorem 1, and applied 100 times the algorithm of Section 4. (We did not run the algorithm for *webdocs* with $w = 2, 3$ due to the inefficiency of the current implementation of the algorithm which does not use the count-min filter.) In all cases we considered, the size suggested by theoretical bound was considerably smaller than the size of the dataset. In particular, for *kosarak* the bound suggested a sample size approximately 20% of the size of the dataset, while for *webdocs* it was between 5% and 10%, and for *T10I4D100K* around 40%, because this last dataset has a smaller number of transactions.

---
[4] http://fimi.cs.helsinki.fi/data/.

| Dataset | #Items | Avg. Trans. Length | # Transactions |
|---------|--------|--------------------|----------------|
| *T10I4D100K* | 1,000 | 10.1 | 100,000 |
| *kosarak* | 41,270 | 8.1 | 990,002 |
| *webdocs* | 5,267,656 | 177 | 1,692,082 |

**Fig. 1.** Datasets characteristics

In all of the runs, for any tested combination of parameters, the output was an $\varepsilon$-approximation to $\text{TOPK}(\mathcal{D},\mathcal{I},K,w)$. This should be compared to the $(1 - \delta) = 0.9$ probability of obtaining a $\varepsilon$-approximation guaranteed by the theorethical results. Also, in all cases the output included at least 95% of the actual top-$K$ frequent itemsets. (Note that the definition of $\varepsilon$-approximation gives no guarantee on the fraction of actual top-$K$ frequent itemsets returned.) In fact, we observed that the actual top-$K$ frequent itemsets discovered from the sample were usually many more than those with actual frequency greater than $f^{(K)} + \varepsilon$, which are guaranteed to be included in the output by definition of $\varepsilon$-approximation. Most of the time (85% of the runs), the output contained exactly all of the actual top-$K$ frequent itemsets. For $w = 2, 3$ frequent itemsets of size greater than one were also correctly identified by mining the sample. In particular, for *kosarak* and $w = 2$, we always correctly identified all the frequent itemsets of size 2 at $k = 5$ (3 such itemsets), $k = 10$ (5 itemsets), and $k = 100$ (65 itemsets). For $k = 1000$ we always identify at least 750 such itemsets of size 2, out of 765. For $w = 3$ we were always able to identify all itemsets of length 2 (4 itemsets) and length 3 (1 itemset) when $k = 10$. For *T10I4D100K*, no frequent itemsets of size greater than 1 existed for the tested values of $k$. Finally, as far as the reported frequency is concerned, in all of the runs and for every itemset, the error between the reported frequency in the output and the real frequency of the itemset was much smaller than $\varepsilon$, usually between $\varepsilon/10$ and $\varepsilon/5$. The higher accuracy of the observed output with respect to what promised by the theoretical analysis is explained by the fact that the latter relies on several approximations which weaken the bounds.

### 6.2 Effectiveness of the stopping condition

Below, we provide both analytical and experimental evidence that the stopping condition used by the algorithm presented in Section 4 is effective in the sense that, for certain datasets, the algorithm stops after mining a sample of size smaller than the upper bound of Theorem 1.

*Analytical evidence.* Consider using the algorithm presented in Section 4 for mining of an $\varepsilon$-approximation to TOPK$(\mathcal{D},\mathcal{I},K,w)$ for a dataset $\mathcal{D}$ over a set $\mathcal{I}$ of $n$ items with confidence at least $1-\delta$. Recall that the algorithm probes increasing sample sizes $t_j$, with $j \geq 0$, until the stopping condition is met or $j = j_{\max}$, where the last sample size $t_{j_{\max}}$ is the minimum between the dataset size and the upper bound given in Theorem 1. For convenience, fix $K = \sum_{i=1}^{w} \binom{\ell}{i}$ for some integer $\ell > w$, and choose the parameters $n, \varepsilon$ and $w$ in such a way to guarantee that $n > \ell$, $j_{\max} > 0$, and $p_K = (h(j)+1)\varepsilon + (\varepsilon/2) + (1/t_j) < 1$, for some $j$ with $0 \leq j < j_{\max}$. It can be easily shown that meaningful configurations of the parameters for which these conditions are satisfied exist (more details will be provided in the full paper). Fix one such value $j$.

Let $\mathcal{I} = \{x_1, x_2, \ldots, x_n\}$ and define $\tau_0 = \{x_1, x_2, \ldots, x_\ell\}$, and $\tau_i = \{x_i\}$, for $\ell < i \leq n$. Consider a dataset $\mathcal{D}$ consisting of $N$ copies of $\tau_0$ and one copy of $\tau_i = \{x_i\}$ for each $\ell < i \leq n$. Thus $\mathcal{D}$ contains a total of $N + n - \ell$ transactions. We allow $N$ to grow arbitrarily large and assume it is large enough to make $N + n - \ell > t_j$ and to make the frequency of each of the $K$ itemsets included in $\tau_0$ greater than $p_K$. We have:

**Theorem 6.** *For a dataset $\mathcal{D}$ built as described above, the algorithm will stop at round $j$ with probability at least $1 - \delta - o(1)$.*

*Proof.* Suppose we are at round $j$ of the algorithm and that Lemma 1 holds, which happens with probability at least $1 - \delta$. Moreover, assume there is no itemset from the $m - K$ not appearing in transaction $t_0$ that has a frequency in the sample greater than $\frac{1}{t_j}$. The probability of this second event is $1 - o(1)$ if $N$ is large enough.

From Lemma 1, we have that for any itemset $x \in B_j(i)$, $0 \leq i \leq h(j)$:

$$f_\mathcal{S}(X) > f_\mathcal{D}(X) - (i+1)\frac{\varepsilon}{2} > f_\mathcal{D}(x) - (h(j)+1)\frac{\varepsilon}{2}.$$

Then, the $K$ itemsets in transaction $t_0$ have frequency in the sample greater than $(h(j)+1)\frac{\varepsilon}{2} + \frac{\varepsilon}{2} + \frac{1}{t_j} > \frac{1}{t_j}$ and are thus the top-$K$ frequent itemsets in the sample. This means they belong to $B_j(0)$, and then $f_\mathcal{S}^{(K)} > p_K - \frac{\varepsilon}{2}$. Hence we have $f_\mathcal{S}^{(K)} - f_\mathcal{S}^{(i)} > (h(j)+1)\varepsilon, \forall i, K < i \leq m$, and the theorem follows. □

*Experimental evidence.* The ability of the stopping condition of halting the sampling schedule before the sample size established by Theorem 1 is reached was also observed when running the algorithm from Section 4 on *kosarak* with the same configurations of parameters $K, w, \varepsilon$ and $\delta$ described in the previous subsection. For this dataset, when $w = 1$ the algorithm always terminated when the sample size was equal to the theoretical bound, while for $w = 2, 3$ it sometimes stopped earlier. Fig. 2(a) and Fig. 2(b) show a comparison between the theoretical bound and the sample size at which the algorithm terminated in our experiments because the stopping condition was satisfied, as function of $K$. We observe that the gap between the stopping sample size and the theoretical bound increases with $w$, which suggests that the stopping condition employed by our algorithm becomes more effective when the number of potential candidate itemsets (i.e., the size of the set $U(\mathcal{I}, w)$) increases.

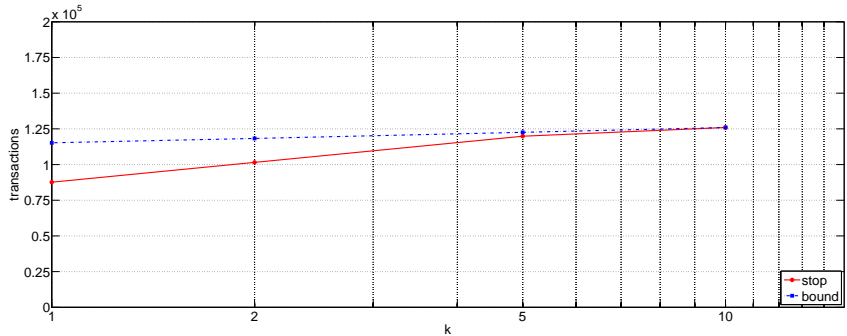

(a) $w = 2$

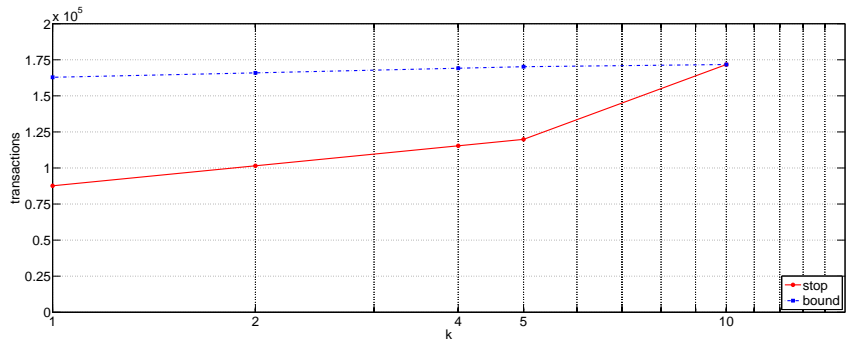

(b) $w = 3$

**Fig. 2.** Experimental evaluation of the effectiveness of the stopping condition for *kosarak*. The dotted line represents the theoretical bound from Theorem 1, the solid line the size at which the algorithm stopped.

## 7 Conclusions

We studied the extraction of the top-$K$ frequent itemsets of bounded size from random samples of a dataset of transactions. We defined a reasonable approximation of the task and explored the tradeoff between the size of the sample and the accuracy of the approximation. In particular, we proved a bound on the sample size sufficient to achieve a given accuracy of the approximation with a given confidence, and we showed that, under certain constraints on the parameters, the bound is tight within constant factors. To the best of our knowledge, this is the first tight relation between sample size and accuracy of the approximation for mining top-$K$ frequent itemsets. We also proposed a progressive sampling algorithm that, in some cases, is able to ensure similar accuracy and confidence while mining smaller samples. For this algorithm, whose efficient implementation is challenging, we proposed an optimization based on count-min filters, a variation of Bloom filters. The effectiveness of our results has been assessed on both artificial and real benchmark datasets. Future research could aim at obtaining tight upper and lower bounds on the sample size required to ensure given accuracy and confidence in all cases, at characterizing the datasets and parameter configurations for which the progressive sampling algorithm becomes profitable, and at engineering an efficient implementation of this algorithm based on the count-min filter.